\documentstyle[aps,twocolumn]{revtex}

\begin{document}
\author{R. E. Prange\cite{Md}}
\title{Quasiclassical Random Matrix Theory}
\address{Institute de Physique Nucl\'eaire\cite{IPN}, Orsay, France \\
Laboratoire de Physique Quantique\cite{PS}, Universit\'e Paul Sabatier,
Toulouse, France}
\date{May 12, 1996}
\maketitle

\begin{abstract}
We directly combine ideas of the quasiclassical approximation with random
matrix theory and apply them to the study of the spectrum, in particular to
the two-level correlator. Bogomolny's transfer operator $T$,
quasiclassically an $N\times N$ unitary matrix, is considered to be a random
matrix. Rather than rejecting all knowledge of the system, except for its
symmetry, [as with Dyson's circular unitary ensemble], we choose an ensemble
which incorporates the knowledge of the shortest periodic orbits, the prime
quasiclassical information bearing on the spectrum. The results largely
agree with expectations but contain novel features differing from other
recent theories.
\end{abstract}


The study of wave systems of low symmetry is known, somewhat misleadingly,
as ``quantum chaos''. Aside from experiment and numerics, the most important
tools in this field are the {\em quasiclassical approximation} [QCA]\cite
{reviews}, and {\em random matrix theory} [RMT]\cite{mehta}.

These two approaches are complementary. Standard RMT treated problems where
knowledge of the Hamiltonian is almost nil. It identifies {\em universal}
features, notably of the {\em spectral statistics}, which statistically
persist for all quantum systems with time reversal invariance, say.

Standard RMT fails for the most fundamental spectral statistic, however,
namely the mean or smoothed [Weyl] density of states, $\bar d(E)$, which is
not, after all, so broadly universal. However, it is argued that {\em local }%
quantities can be calculated, i.e. quantities involving energies in the
range $E\pm W$, where $W$ $\bar d(E)>>1,$ and $W$ is small enough so that $%
\bar d(E)$ can be taken as constant. RMT treats the {\em Hamiltonian,} $%
{\cal H},$ as a random $N\times N\,$ matrix, locally normalizes to $\bar d$,
and lets $N$ $\rightarrow \infty .$

In the QCA, a system is chosen, e.g. a particular stadium billiard, and the
smallness of the ratio [wavelength]/[characteristic classical length] is
exploited. A particular spectrum, not just statistics, is in principle
obtained. The basic idea, of Gutzwiller\cite{GTF} but much advanced by others%
\cite{voroszeta},\cite{berrykeat},\cite{bogolss},\cite{dorsmil}, makes two
subtly related but ostensibly different QCA approximations.

The first is the Weyl approximation to $\bar d$ whose leading order is the
phase space integral $\bar d(E)=\frac 1{(2\pi \hbar )^f}\int
d^fpd^fq\,\delta (E-{\cal H}(p,q))$ where the dimension $f$ is here taken to
be $2.$ The density of states is divided into $d(E)\equiv \sum \delta
(E-E_a)=\bar d(E)+d_{osc}(E)$. A WKB-like approximation is made to give $%
d_{osc}$, describing the oscillations about the Weyl term, $%
d_{osc}(E)=\sum_pA_p\cos S_p(E)/\hbar .$ The sum is over {\em periodic
orbits }of action $S_p$ [which includes the Maslov index] and $A_p$ depends
on the classical stability of the orbit. In this primitive form, the result
is the {\em Gutzwiller trace formula}\cite{GTF}.

Many advances have been made, e.g. $d_{osc}$ can be written as a ratio of
absolutely convergent series, rather than as the divergent series above\cite
{gp1}. The major result is that by such means only orbits whose period $%
T_p=\partial S_p/\partial E$ is less than {\em half }the ``Heisenberg
time'', $\tau _\hbar =2\pi \hbar \bar d$ $\propto 1/\hbar $ need be
considered: All longer orbits can be effectively expressed in terms of
these. Orbits longer than $\tau _\hbar $ combine to give nothing, those
shorter display a sort of reflection symmetry [resurgence] about $\frac 12%
\tau _\hbar .$

No direct way to solve a fundamental problem is known, however. Namely, the
number of periodic orbits {\em proliferates exponentially} with $T_p$ and
for small $\hbar $ no one can actually find such long orbits, let alone add
them up. Therefore recourse is made to a {\em statistical treatment} of the
long orbits, via the Hannay-Ozorio de Almeida sum rule\cite{HanOA}, valid
for hard chaotic systems. This requires a form of the{\em \ diagonal
approximation }[DA]. [The product $d_{osc}d_{osc}$ is expanded in a double
Gutzwiller series, and only the diagonal terms are kept. After the DA,
quantum effects enter only trivially through the energy scale $\bar d.$] The
DA by itself is not sufficient, and it fails dramatically [at long time]
since long orbits need to show important correlations, if they are to
describe a discrete spectrum, and give results in agreement with RMT and
experiment for neighboring level properties\cite{offdiag}. [ A recent effort
of Bogomolny and Keating\cite{BK} builds in the discreteness of the spectrum
in an ad hoc but effective way. The supersymmetry [SS] technique has begun
to find good results in this connection also\cite{AAA}.] In short, the QCA
has to invoke statistics to deal with the long orbits. On the other hand,
there are many influences of the short periodic orbits on the spectrum which
are observable and have been of great interest.

A crucial advance in semiclassics was made by Bogomolny\cite{bogolss} with
his $T$ operator. The $T$ operator is an approximation to and/or
generalization of the {\em boundary integral method } used to study billiard
problems. It is a QCA to an exact kernel $K(q,q^{\prime },E)$ [where $%
q,q^{\prime }$ run over the boundary of the billiard] such that the spectrum
is given by the zeroes of $\det (1-K(E)).$ The equivalence of the
`resurgence' mentioned above was shown: Namely, in QCA $T$ is {\em unitary}
and {\em of} {\em rank} $N$, i.e. equivalent to an $N\times N$ unitary
matrix $T_{mn}(E),$ Further, a general Poincar\'e {\em surface of section}
can be used instead of the boundary, extending the technique beyond
billiards.

{\bf Our proposal }is to treat the matrix $T_{mn}$ as a {\em random} unitary
matrix, generalizing Dyson's famous work\cite{mehta}. This has several
advantages over the usual choice of the Hamiltonian as random matrix.

\begin{itemize}
\item  The mean distribution of eigenvalues $e^{i\theta _n}$, $\bar d(\theta
),$ is physically meaningful, and related to $\bar d(E).$ [We distinguish
the distributions in angle and energy by the argument. We also denote, by $%
\bar d$ without an argument, the Weyl density of states. Dyson's COE and CUE
matrix ensembles have $\bar d(\theta )=N/2\pi .$] The study of deviations
from $\bar d$ constant is the purpose of $d_{osc}$. Random Hamiltonian
matrices are unable to deal with such questions.

\item  The rank $N$ is meaningful. It is [usually] given by $N\simeq
2L/\lambda (E)$ where $L$ is the length of the surface of section, and $%
\lambda $ is the wavelength at energy $E.$

\item  The matrix size is $N$ $\propto \hbar ^{-1}$ rather than $\hbar ^{-2}$%
, an important efficiency of the boundary integral method.

\item  {\em Specific quasiclassical knowledge} may be incorporated into the
definition of the ensemble of matrices in a natural and convenient way. In
particular, one may study modified Dysonian ensembles, which have {\em %
prescribed mean values of the traces} $\sigma _r=%
\mathop{\rm Tr}
T^r.$ The set of traces $\sigma _{r\text{, }}r=1..N$ uniquely determine $%
\det (1-T)$ and thus the spectrum. Low order traces are usually well
evaluated quasiclassically by the method of stationary phase. They are
expressed in terms of {\em short periodic orbits, }and we take this
approximation for $\sigma _r$. The high order traces cannot usually be so
evaluated as the number of orbits is too large.
\end{itemize}

Some difficulties with the approach are

\begin{itemize}
\item  Direct calculation gives the eigenphases $\theta _n$ rather than the
energies $E_{a\text{ }}$ for which $\theta _n\left( E_a\right) =0%
\mathop{\rm mod}
2\pi $. However, there is evidence that the $\theta _{n\text{ }}$are locally
linear functions of the energy. For $N$ large, this is sufficient, as Dyson
showed. [Bogomolny, in unpublished lectures, has made this connection more
precise.] For billiards one can probably do better, since scaling indicates
that $\theta _n=a(k-k_n)$ where $k=\sqrt{E}$. [We put $\hbar =1.$]

\item  We have assumed that $N$ is an integer which depends on energy.
Therefore, rather unphysical jumps in $N$ are needed as the energy
increases. Perhaps the simplest way around this is to express the results as
a function of $N$ and `analytically continue' to continuous $N$.
\end{itemize}

Let $Z(E)=\det (1-T)=\prod (1-e^{i\theta _n}).$ Then 
\begin{eqnarray}
d_{osc} &=&%
{\textstyle {-1 \over ^\pi }}
\mathop{\rm Im}
{\textstyle {d \over dE}}
\ln Z(E)=-%
{\textstyle {1 \over 2\pi }}
\sum_n\theta _n^{\prime }\left( 1+%
\mathop{\rm Im}
\cot 
{\textstyle {1 \over 2}}
\theta _n\right)  \nonumber  \label{dosc2} \\
\ &=&\sum_n\left[ 
{\textstyle {-1 \over 2\pi }}
\theta _n^{\prime }+\theta _n^{\prime }\delta (\theta _n(E))\right]
\label{dosc2}
\end{eqnarray}
Assuming $\theta _n(E)=(E-E_n)\theta _n^{\prime }$ with $\theta _n^{\prime }$
$\approx \theta ^{\prime }$ approximately independent of $n$, gives $\bar d=$
$\bar d_{Weyl}(E)=\frac 1{2\pi }\sum_n\theta _n^{\prime }=\frac N{2\pi }%
\theta ^{\prime }.$ For a billiard, $\theta ^{\prime }=a/2k$. Using the
estimate for $N$ it is found that $\bar d=La/4\pi ^2.$ Thus $a=A\pi /L,$
where $A\,$ is the area of the billiard.

We now add knowledge of the short period orbits {\em and only this knowledge}
to the problem, in the spirit of the information theoretical approach of
Balian\cite{balian}. We cite other recent work using this technique\cite
{mutalib}. Let $d\mu (T)$ be the measure for the class of random matrices
under consideration. We want the probability distribution $P(T)d\mu (T)$.
Then the `entropy' $S[P]=-\int d\mu (T)P(T)\ln (P(T))$ is maximized, subject
to conditions $\int P(T)%
\mathop{\rm Tr}
T^rd\mu (T)=\sigma _r;$ $\,r=1,2,..$

For simplicity we choose the measure without time reversal symmetry
corresponding to Dyson's CUE. $T$ can be parametrized by its $N$ eigenphases 
$\theta _n$ together with $N(N-1)$ other real parameters $b_s$, which we
integrate out. Then $d\mu (T)=\prod_{n<m}\left| e^{i\theta _m}-e^{i\theta
_n}\right| ^2\prod_nd\theta _n.$ Introducing Lagrange multipliers for the
constraints, we obtain 
\begin{equation}
p(\theta _1,\theta _2,..\theta _N)\prod_nd\theta _n=\prod_{r<s}\left|
e^{i\theta _s}-e^{i\theta _r}\right| ^2\prod_nw(\theta _n)d\theta _n
\label{pdist}
\end{equation}
where the positive weight factor is 
\begin{equation}
w(\theta )=\exp \left[ \sum \lambda _r\cos r(\theta -\vartheta _r)\right] .
\label{w0}
\end{equation}
The $\lambda $'s and $\vartheta $'s are the Lagrange multiplier parameters
[two for each complex $\sigma _r]$ needed to specify the traces and $\lambda
_r$ vanishes for an unconstrained trace. We study $\bar d(\theta ),$ $%
R(\theta _1,\theta _2)$ obtained by integrating out all but one or two $%
\theta $'s in Eq.(\ref{pdist}).

The method of orthogonal polynomials is used\cite{mehta},\cite{szego}.
Polynomials $\phi _n(z)\,$in $z=e^{i\theta }$ are found, satisfying $\frac 1{%
2\pi }\int d\theta w\left( \theta \right) \phi _n\left( z\right) \overline{%
\phi _m(z)}=\delta _{nm}$ where the bar indicates complex conjugation. This
and subsequent integrals are over $[-\pi ,\pi ]$. Eq.(\ref{pdist}) can be
expressed $p(\theta _1,..\theta _N)=\det M^{\dagger }M=\det K$ where $\sqrt{%
2\pi }M_{nm}=\Phi _n(\theta _m)=\sqrt{w(\theta _m)}\phi _n(\theta _m)$ and $%
K_{ij}=K(\theta _i,\theta _j)=\frac 1{2\pi }\sum_{n=0}^{N-1}\Phi _n(\theta
_i)\overline{\Phi _n(\theta _j)}$ . The reason for writing $p$ in this form
is that an integral over one variable of an $R\times R$ determinant of the
form $\det K$ is proportional to the $(R-1)\times (R-1)$ determinant over
the same argument. We thus find that 
\begin{equation}
\bar d(\theta )=K(\theta ,\theta )  \label{d3}
\end{equation}
and 
\begin{equation}
R(\theta _1,\theta _2)=\bar d(\theta _1)\bar d(\theta _2)-\left| K(\theta
_1,\theta _2)\right| ^2  \label{R2}
\end{equation}

The sum in the definition of $K$ can be carried out\cite{szego}. The formula
[Christoffel-Darboux] is 
\begin{equation}
\sum_{n=0}^{N-1}\phi _n(z_1)\overline{\phi _n(z_2)}=\frac{^{z_1^N\bar \phi
_N(z_1^{-1})\bar z_2^N\phi _N(\bar z_2^{-1})-\phi _N(z_1)\overline{\phi
_N(z_2)}}}{1-z_1\bar z_2}.  \label{C-D}
\end{equation}
[$\bar \phi (z)\equiv \overline{\phi (\bar z)}.$] We thus express everything
in terms of the single polynomial $\phi _N$ [and $w$].

Given $w(\theta ),$ satisfying certain conditions, a formula exists for $%
\phi _N$ for $N\rightarrow \infty .$ The weight function of Eq.(\ref{w0}) is
particularly congenial. If the exponent is a convergent series, the
conditions are well satisfied.

The weight can be uniquely expressed by $w(\theta )=\left| D(z)\right| ^2$.
Ambiguity in $D$ is removed by insisting that $D(z)\neq 0$ for $\left|
z\right| <1$, and $D(0)=1$. For Eq.(\ref{w0}), $D(z)=\exp \left[ 
{\textstyle {1 \over 2}}
\sum \lambda _ne^{-in\vartheta _n}z^n\right] .$ Asymptotically, for large $N$%
, the theorem\cite{szego} is that 
\begin{equation}
\phi _N(z)\approx z^N\left[ \bar D(z^{-1})\right] ^{-1}  \label{phiN}
\end{equation}
Assuming asymptotia has been reached, we use Eq.(\ref{phiN}) in Eq.(\ref{C-D}%
) and Eq.(\ref{d3}) to find, with L'H\^opital's rule, that 
\begin{equation}
\bar d(\theta )\approx \frac N{2\pi }+\frac 1\pi 
\mathop{\rm Re}
\left[ z\frac{D^{\prime }(z)}{D\left( z\right) }\right] .  \label{d4}
\end{equation}
The integral over the nontrivial term vanishes. Thus 
\begin{equation}
\bar d(\theta )=\frac N{2\pi }+\frac 1{2\pi }\sum n\lambda _n\cos n(\theta
-\vartheta _n).  \label{d5}
\end{equation}

The trace conditions are $\sigma _r=\int d\theta \,e^{ir\theta }\bar d%
(\theta )=%
{\textstyle {1 \over 2}}
r\lambda _re^{ir\vartheta _r}$, determining the Lagrange parameters. The
mean density of states in energy is $\bar d(E)=\theta ^{\prime }\bar d%
(\theta =0)=\bar d+\frac{\theta ^{\prime }}\pi \sum \left| \sigma _r\right|
\cos (r\vartheta _r)$. Gutzwiller's density of states, expressed in terms of
the traces of $T$ is 
\begin{eqnarray}
d_{Gutz}(E) &=&\bar d+%
{\textstyle {1 \over \pi}}
\mathop{\rm Im}
{\textstyle {d \over dE}}
\sum 
{\textstyle {1 \over r}}
\sigma _r(E)  \nonumber  \label{d7} \\
\ &\approx &\bar d+%
{\textstyle {1 \over \pi}}
\sum \vartheta _r^{\prime }\left| \sigma _r\right| \cos (r\vartheta _r)
\label{d7}
\end{eqnarray}
where for pedagogical purposes we assume that $e^{ir\vartheta _r}$ is
rapidly varying. If the derivative $\vartheta _r^{\prime }$ $\approx $ $%
\theta ^{\prime }$ , the leading terms of $d_{Gutz}$ coincide with the RMT $%
\bar d(E).$

We thus have the interpretation that, at the level of the mean density, the
random $T$ operator scheme is equivalent to moving the long wavelength part
of $d_{osc}$ into the mean density of states. {\em The formulas above give a
way of including this knowledge into a prediction of the correlation
functions.}

We next evaluate the two point correlations, using Eq.(\ref{R2}). Define $%
\psi (\theta )[=\frac 12\sum \lambda _n\sin n(\theta -\vartheta _n)]$ by $%
e^{i\psi (\theta )}=D(z)/\left| D(z)\right| .$ [For Dyson, $\psi =0.$] Then 
\[
\psi (\theta _1)-\psi (\theta _2)=\sum \lambda _n\sin 
{\textstyle {n \over 2}}
(\theta _1-\theta _2)\cos 
{\textstyle {n \over 2}}
(\theta _1+\theta _2-2\vartheta _n) 
\]
\begin{equation}
\left| K(\theta _1,\theta _2)\right| ^2=\left[ \frac{\sin \left[ \frac N2%
(\theta _1-\theta _2)+\psi (\theta _1)-\psi (\theta _2))\right] }{2\pi \sin 
\frac 12(\theta _1-\theta _2)}\right] ^2  \label{Ksqr}
\end{equation}
agreeing with Dyson if $\psi =0.$ We wish to calculate 
\begin{eqnarray}
C(x) &=&\left\langle C_\epsilon (x)\right\rangle _W=\left\langle
\left\langle 
{\textstyle {1 \over \bar d^2}}
d(\epsilon +%
{\textstyle {x \over 2\bar d}}
)d(\epsilon -%
{\textstyle {x \over 2\bar d}}
)\right\rangle \right\rangle _W  \nonumber  \label{Cofx} \\
\ &=&\left\langle \left\langle 
{\textstyle {1 \over \bar d^2}}
\sum_{a,b}\delta (\epsilon +%
{\textstyle {x \over 2\bar d}}
-E_a)\delta (\epsilon -%
{\textstyle {x \over 2\bar d}}
-E_b)\right\rangle \right\rangle _W  \nonumber  \label{C(x)} \\
&=&\left\langle R(\epsilon +%
{\textstyle {x \over 2\bar d}}
,\epsilon -%
{\textstyle {x \over 2\bar d}}
)+%
{\textstyle {1 \over \bar d}}
\delta (%
{\textstyle {x \over \bar d}}
)\right\rangle _W  \label{C(x)}
\end{eqnarray}
The outer average is over the energy range $E\pm W.$ The inner average is
over the random matrix ensemble.

We assume as above that $\theta _a(E)=(E-E_a)\theta ^{\prime }$ . Then 
\begin{equation}
R(E_1,E_2)=(%
{\textstyle {\theta ^{\prime } \over \bar d}}
)^2R\left( \left( E-E_1\right) \theta ^{\prime },\left( E-E_2\right) \theta
^{\prime }\right)  \label{R(E)}
\end{equation}

Using this in Eq.(\ref{C(x)}), with $\epsilon =E$ we obtain 
\begin{eqnarray}
C_\epsilon (x) &=&\frac 1{\bar d}\delta (\frac x{\bar d})+\left( \frac{%
\theta ^{\prime }}{\bar d}\right) ^2\bar d\left( -\frac{\theta ^{\prime }x}{2%
\bar d}\right) \bar d\left( \frac{\theta ^{\prime }x}{2\bar d}\right) - 
\nonumber  \label{C2(x)} \\
&&\left[ \frac{\sin \left[ \frac{\theta ^{\prime }Nx}{2\bar d}+\sum \lambda
_n\sin \frac{n\theta ^{\prime }x}{2\bar d}\cos n\vartheta _n\right] }{\frac{%
2\pi \bar d}{\theta ^{\prime }}\sin \frac{\theta ^{\prime }x}{2\bar d}}%
\right] ^2  \label{C2(x)}
\end{eqnarray}
Since $\theta ^{\prime }/\bar d=2\pi /N$ is small, we find (suppressing $%
\delta (x)$) 
\begin{eqnarray}
C_\epsilon (x) &=&\left( 1+\sum c_{n+}\right) \left( 1+\sum c_{m-}\right) 
\nonumber  \label{C3(x)} \\
&&\ \ \ \ \ -\left[ \frac{\sin \pi x\left( 1+\sum \frac{n\lambda _n}N\cos
n\vartheta _n\right) }{\pi x}\right] ^2  \label{C3(x)}
\end{eqnarray}
where $c_{n\pm }(x)=\frac{n\lambda _n}N\cos n(\frac{\theta ^{\prime }x}{2%
\bar d}\pm \vartheta _n)$

We now perform the energy average. Assume that only the $\vartheta _n$'s
vary significantly, and that they can be averaged independently. This gives
the DA for the squared terms $c_{n+}c_{n-\text{ }}$in the first line of Eq.(%
\ref{C3(x)}), equivalent to that found by Berry\cite{BFF}. We call this term 
$C_B(x)=\frac 12\sum \left( \frac{n\lambda _n}N\right) ^2\cos \frac{2\pi n}N%
x.$ Let $\rho (y)=\rho (-y)=\left\langle \delta (y-\sum \frac{n\lambda _n}N%
\cos n\vartheta _n)\right\rangle _W$ be the distribution of the correction
term in the second line of Eq.(\ref{C3(x)}). Its width is of order $N^{-1}.$
The form factor\cite{BFF} is 
\begin{eqnarray}
k(t) &\equiv &1+\int dxe^{2\pi itx}C(x)=\sum \left( \frac{n\lambda _n}{2N}%
\right) ^2\delta (\left| t\right| -t_n)  \nonumber  \label{ff} \\
\;\;\;\;\,\,\,\,\,\,\,\,\,\,\,\,\,\,\ &&+\left( 1-\int_{\left| t\right|
-1}^\infty dy\int_y^\infty dy^{\prime }\rho (y^{\prime })\,\right)
\label{ff}
\end{eqnarray}
where $t_n\left( =n/N\right) $ is the period of the orbits associated with
the $n$'th trace. The first term is a simulation of Berry's DA result for
the short orbits. The second term rounds off the sharp corner found in
standard RMT, [where $\rho =\delta (y)$.] The magnitude and width of the
roundoff are both $\sim N^{-1}$. Since $\rho \geq 0,$ we do not find the
suggested\cite{AAA} oscillations in $k(t)$ near $t=1$.

This expression mostly agrees with what is believed to be the correct form
of $k(t).$ For large $\left| t\right| $, the discreteness of the spectrum
dominates. For $\left| t\right| \sim t_c\sim N^{-1}$, the DA applied to
Gutzwiller's expression gives spikes when $\left| t\right| $ is at a period
of a short periodic orbit. We should improve our elementary estimate to
replace $C_B$ as given above by Berry's result, and we should also make the
corresponding change in $\psi (\theta )$. For $t_c<t<<1,$ the DA applies and
the sum rule\cite{HanOA} gives a straight line of unit slope. There is then
structure near $\left| t\right| =1$ which is a consequence of the short
period structure. The area removed under $k(t)$ near $t=1$ just compensates
the area gained from the spikes at $t\sim t_c,$ a consequence of $R(\theta
,\theta )=0$. It's unclear whether resurgence plays a role in producing
structure near $\left| t\right| =1$, as has been suggested\cite{AAA}. Our
formulation contains all the ingredients needed for resurgence. RMT
fluctuations in $\det T$ may wash out this effect. It seems rather that the
energy average produces its biggest effect at $\left| t\right| =1$ because
of the kink there.

The replacement $\sin \pi nx/N\rightarrow \pi nx/N$ in obtaining Eq.(\ref
{C3(x)}) from Eq.(\ref{C2(x)}) is not necessary. Averaging as before, we
find 
\begin{eqnarray}
&&\ \ \left\langle \cos \left( 2\pi x+2\sum \lambda _n\sin \frac{\pi nx}N%
\cos n\vartheta _n\right) \right\rangle _W  \nonumber  \label{AveCos} \\
\  &\approx &\cos 2\pi x\prod \left( 1-\lambda _n^2\sin ^2\frac{\pi nx}N%
\right) \equiv x^2{\cal D}(x)\cos 2\pi x  \label{D(x)}
\end{eqnarray}
and we have assumed $\lambda _n\ $small. Then 
\begin{equation}
C(x)=\frac 1{2\pi ^2}\left( -\frac 12\frac{\partial ^2\ln {\cal D}(x)}{%
\partial x^2}+{\cal D}(x)\cos 2\pi x\right) ,  \label{CD}
\end{equation}
the relation found by SS\cite{AAA}. With Eq.(\ref{D(x)}) small structures
near $\left| t\right| =1$ decorate the result of Eq.(\ref{ff}) but the
result is still monotonic.

It is not clear under what conditions Eq.(\ref{CD}) holds. The SS result is
derived for large $x,$ while ours is for small $\lambda _n$, in other words,
for rather unstable orbits. The ${\cal D}(x)$ corresponding to a
distribution $\rho $, $\,$is ${\cal D}(x)=x^{-2}\int dy\rho (y)\cos 2\pi xy$%
, but the term with $\ln {\cal D}$ does not reproduce $C_B$. The ${\cal D}$
of \cite{AAA} is derived from $C_B$ as a correction to the sum rule: it
corresponds to a non-positive `distribution' $\rho (y)\propto \cos 2\pi
\gamma _2y\exp \left( -2\pi \gamma _1\left| y\right| \right) $ where $\gamma
_1+i\gamma _2$ is the lowest nonvanishing eigenvalue of the Perron-Frobenius
operator. It of course does not give the spiky structure of $k(t)$ at $t\sim
t_c.$

To summarize, an application of random matrix theory to Bogomolny's $T$
operator, one which retains knowledge of the short periodic orbits, leads to
a relatively simple theory which gives reasonable results for the density of
states and for the two point correlations. It does not agree in every detail
with previously obtained results, however. This may be due to approximations
we have made as well as the fact we have not specifically considered orbits
of length of order $\gamma _1^{-1}>t_c.$

In this note, we have not worked out what happens for finite $N$, although
the results should be meaningful. One can find the results for higher
correlations directly. We don't expect them, or the corrections to the
nearest neighbor level spacing, to be very illuminating, however. We hope to
extend the results to other symmetry classes, and to parametric
correlations. It should also be possible to compare statistics of
eigenphases directly with this theory, in the case of quantum maps, [and
indeed the $T$ operator is the quantum Poincar\'e map.]

We thank O. Bohigas and P. Leb\oe uf for extensive discussion and for
hospitality at the INP, and E. Bogomolny for informing us of his unpublished
work. We thank J. Bellissard for hospitality at the Universit\'e Paul
Sabatier.

\end{document}